\documentclass[a4paper]{ISOStyle}
\usepackage{epsfig}

\begin{document}

\title{Photometric Mapping with ISOPHOT using the
``P32'' Astronomical Observation Template}

\author{R.\,J.\,Tuffs\inst{1} \and C.\,Gabriel\inst{2}}
  \institute{Astrophysics Division, Max-Planck-Institut f\"ur Kernphysik,
     Saupfercheckweg 1, 69117 Heidelberg, Germany. 
     Richard.Tuffs@mpi-hd.mpg.de
  \and
     ISO Data Centre, Astrophysics Division,
     ESA, Villafranca del Castillo, P.O. Box 50727, 28080 Madrid, Spain
}

\maketitle 

\begin{abstract}

The ``P32'' Astronomical Observation Template (AOT) provided a means to
map large areas of sky (up to 45$\times$45 arcmin) in the FIR at 
high redundancy and with sampling close to the Nyquist limit 
using the ISOPHOT C100 (3$\times$3) and C200 (2$\times$2)
detector arrays on board ISO. 
However, the transient response behaviour of the Ga:Ge detectors,
if uncorrected, can lead to severe systematic photometric errors and 
distortions of source morphology on maps.
Here we describe the basic concepts of an algorithm which can
successfully correct for transient response artifacts in P32 observations. 
Examples are given demonstrating the photometric and imaging performance 
of ISOPHOT P32 observations corrected using the algorithm. 

\keywords{ISO}

\end{abstract}

\section{INTRODUCTION}
From the point of view of signal processing and photometry
diffraction-limited mapping in the FIR with cryogenic space 
observatories equipped with photoconductor 
detectors poses a particular challenge. In this wavelength regime 
the number of pixels in detector arrays is limited in comparison with 
that in mid- and near-IR detectors. This means that more repointings 
needed to map structures spanning a given number of 
resolution elements. Due to the logistical constraints imposed
by the limited operational lifetime of a cryogenic mission, this
inevitably leads to the problem that the time scale for modulation of 
illumination on the detector pixels becomes smaller than 
the characteristic transient response timescale of the detectors to 
steps in illumination. The latter timescale can reach minutes.

Unless corrected for, the transient response behaviour of the detectors 
will lead to distortions in images, as well as to systematic errors in
the photometry of discrete sources appearing on the maps. In general,
these artifacts become more severe and more difficult to correct for
at fainter levels of illumination, since the transient response timescales
increase with decreasing illumination. Compared to the IRAS detectors,
the ISOPHOT-C detectors (Lemke et al. 1996) on board the 
Infrared Space Observatory (ISO; Kessler et al. 1996)
had relatively small pixels designed to 
provide near diffraction limiting imaging. ISOPHOT thus generally
encountered larger contrasts in illumination between source and 
background than IRAS did, making the artifacts from the transient 
response more prominent, particularly for fields with faint backgrounds.

A further difficulty specific to mapping in the FIR with ISO was
that, unlike IRAS, the satellite had no possibility to cover a target 
field in a controlled raster slew mode. This limited the field size 
that could be mapped using the spacecraft raster pointing mode alone, 
since the minimum time interval between the satellite fine pointings used
in this mode was around 8\,s. This often greatly exceeded the nominal 
exposure time needed to reach a required level of sensitivity 
(or even for many fields the confusion limit).
Furthermore, the angular sampling and redundancy achievable using the
fine pointing mode in the available time was often quite limited, 
so that compromises sometimes had to be made to adequately extend 
the map onto the background.

A specific operational mode for ISO - the ``P32'' Astronomical 
Observation Template (AOT) - was developed 
for the ISOPHOT instrument to alleviate 
these effects (Heinrichsen et al. 1997). 
This mode employed a combination of standard 
spacecraft repointings and rapid oversampled scans 
using the focal plane chopper. The technique could 
achieve a Nyquist sampling on map areas of sky ranging up to 
45$\times$45 
arcmin in extent (ca. 70$\times$70 FWHM resolution elements) on timescales 
of no more than a few hours. In addition to mapping large sources, the 
P32 AOT was extensively used to observe very faint compact 
sources where the improved sky sampling and redundancy alleviated 
the effects of confusion and glitching.

In all, over 6$\%$ of the observing time of ISO
was devoted to P32 observations during the 1995-1998 mission, 
but the mode could not until now be fully exploited scientifically
due to the lack of a means of correcting for the complex non-linear 
response behaviour of the Ge:Ga detectors. Here we
describe the basic concept of a new algorithm which can
successfully correct for the transient response artifacts in 
P32 observations. This algorithm forms the kernel of the ``P32TOOLS'' 
package, which is now publically available as
part of the ISOPHOT Interactive Analysis package PIA (Gabriel et. al 1997;
Gabriel \& Acosta-Pulido 1999).
Information on the algorithm, as well as the first scientific
applications, can also be found in Tuffs et al. (2002). 
The user interface of P32TOOLS is described by Lu et al. (this volume).
After a brief overview of relevant aspects of the P32 AOT in Sect.~2,
we describe the semi-empirical model used to reproduce the transient 
response behaviour of the PHT-C detectors in Sect.~3. Sect.~4 
describes the algorithms used to correct data. Examples demonstrating
the photometric and imaging performance of ISOPHOT P32 observations
are given in Sect.~5, based on maps corrected using P32TOOLS.

\section{THE P32 AOT}

The basic concept of the P32 AOT, in which imaging was
achieved using a combination of standard spacecraft repointings 
and rapid oversampled scans using ISOPHOT's focal plane chopper, 
is summarised in the contribution by C. Gabriel in this volume. 
There were two basic requirements, outlined in the original proposal 
for the AOT (Tuffs \& Chini 1990) which led to the final design:

The first goal was to give ISOPHOT the capability of
achieving a Nyquist sky sampling 
($\Delta\Theta\,$=$\,\lambda$/2D$\,$=$\,$17$\,$arcsec at 100$\,\rm \mu m$)
on extended sources subtending up to 50 resolution elements
in a tractable observation time. As well as providing a unique 
representation of any arbitrary sky brightnesss distribution, this 
also improved the effective confusion limit in deep observations, 
which, in directions of low background was $\sim100$ 
and 20\,mJy rms at 200 and 100$\,\rm \mu m$, respectively, 
for ISOPHOT-C. 

The second requirement
was to achieve a redundancy in the observed data. This led to the
sampling of each of a given set of sky directions by each detector pixel 
at different times. The redundancy was achieved in part
by making several (at least 4) chopper sweeps at each spacecraft
pointing direction. Further redundancy was achieved by having 
an overlap in sky coverage of at least 
half the chopper sweep amplitude between chopper sweeps made at successive 
spacecraft pointing directions in the Y spacecraft coordinate. Thus,
a given sky direction was sampled on three different time scales
by each detector pixel: on intervals of the detector non-destructive
read interval, on intervals of the chopper sweep, and on intervals
of the spacecraft raster pointings. Although the redundancy
requirement was originally made to help follow the long term trends 
in detector responsivity during an observation, in practice it also 
proved very useful in removing effects caused by cosmic particle
impacts (Gabriel \& Acosta-Pulido 2000), 
the so-called ``glitches'', from the data. This was especially 
important to optimise the sensitivity of deep observations using the 
C100 detector, which are generally limited by glitching rather than 
by confusion.

Both these requirements led to dwell times per chopper plateau
which could be as low as 0.1\,s. For a given sky brightness, this also
had the consequence that the detector was generally read faster than in the 
other AOTs using the PHT-C detectors.

\subsection{The P32 ``natural grid''}

The fact that the spacecraft pointing increment in Y was
constrained to be a multiple of the chopper step interval
means that a rectangular grid of directions on the sky can be
defined such that all data taken during P32
measurement fine pointings will, to within 
the precision of the fine pointings,
exactly fall onto the directions of the grid. This so-called
P32 ``natural grid'' allows maps to be made without any gridding 
function, thus optimising angular resolution. The P32 natural
grid is also a basic building block of the algorithm to correct for the 
transient response behaviour of the detector, as the algorithm solves
for the intrinsic sky brightness in M$\times$N independent variables
for a grid dimension of M$\times$N. An example of the
almost perfect registration of data onto the grid is shown in
Fig.~1.

\begin{figure}[ht]
  \begin{center}
    \epsfig{file=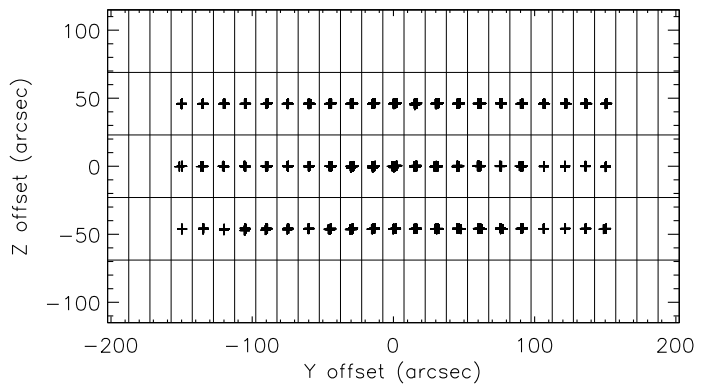, width=5cm, angle=90.}
  \end{center}
\caption{Pointing directions observed towards Ceres
at 105$\,{\mu}m$. The spacecraft raster dimension was $3\,\times\,3$ in 
the spacecraft coordinates $Y\,\times\,Z$. Sky sampling in $Y$ is
achieved through a combination of chopper scans in $Y$ and 
repointings of the spacecraft. In $Z$ the sky sampling is determined
alone by the spacecraft repointing interval. Each pointing direction 
for 11833 individual data samples for the central pixel of the C100 
detector array is plotted as a cross. The rectangular grid
is the ``P32 natural grid'' (see text) for this observation, on which
the sky brightness distribution is to be solved. The grid sampling
is $14.99\times45.98$ arcsec. Only points with the on target flag set
(i.e. not including slews) are plotted.
The pointings typically lie within 1 arcsec of the centre of 
each pixel of the P32 natural
grid.}
\end{figure}

All the example maps shown in this paper will be sampled without
a gridding function on the P32 natural grid. This means that the
data in each of the map pixels is independent. This is unlike 
the situation for most other maps calculated using PIA, where a
gridding function is employed and so raw data can contribute to 
more than one map pixel.

\section{TRANSIENT RESPONSE OF THE PHT-C DETECTORS}

Gallium doped germanium photoconductor detectors exhibit a number of effects
under low background conditions which create severe calibration
uncertainties. The ISOPHOT C200 (Ge:Ga stressed) and especially the C100
(Ge:Ga) detectors have a complex non-linear response
as a function of illumination history on timescales of $\sim0.1-100$\,sec, 
which depends on the absolute illumination as well as
the changes in illumination. The following behaviour is seen 
in response to illumination steps:

\begin{itemize}

\item a rapid jump in signal, which however undershoots
the asymptotic equilibrium response (often severely for the C100 array)

\item a hook response (overshooting or undershooting on intermediate
timescales of $\sim1-10$\,s)

\item a slow convergence to the asymptotic equilibrium response of 
the detector on timescales of $\sim10-100$\,s

\end{itemize}
An example showing these effects for a staring observation
is given in Fig.~2. Also seen are points discarded by the deglitching
algorithm in PIA, marked by stars. Many glitches are 
accompanied by a longer lived ``tail''.
It is these tails that determine the
ultimate sensitivity of the ISOPHOT C100 array.

\begin{figure}[ht]
  \begin{center}
    \epsfig{file=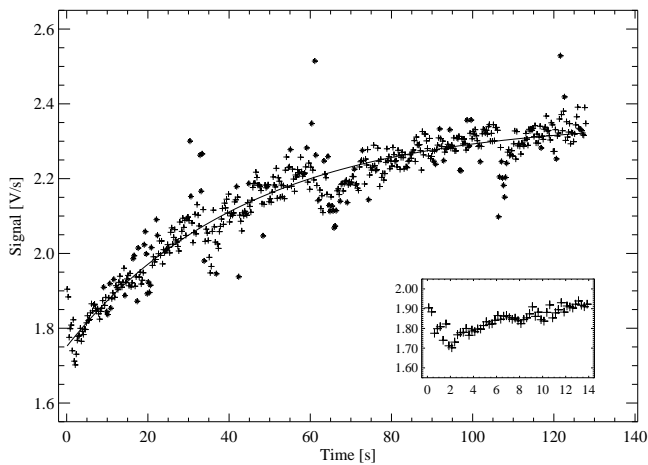, width=8.5cm}
  \end{center}
\caption{Example of the transient response of a Ge:Ga detector
following an upwards step to a constant level of illumination
at 100$\,\rm \mu m$. The hook response is shown in the inset.}
\end{figure}

\subsection{Semi-empirical detector model} 

Unlike the ISOCAM and ISOPHOT-S Si:Ge detectors, the transient
response behaviour of the Ge:Ga detectors of ISOPHOT-C could not be 
described by the Fouks Schubert model (Schubert et al. 1995; 
Acosta-Pulido 1998). However, the effects observed in ISOPHOT-C
can to some extent be qualitatively understood 
in terms of the build up of charge on the contacts between 
semiconductor and readout circuit, as modelled by Sclar (1984).
Sclar's model can be formulated as the superposition of three 
exponentials to describe the response of the detector to a step in
illumination. The three timescales involved are a minimum requirement to
reproduce the hook response and the long term approach to the 
asymptotic response. However, no theory exists to predict values 
for the constants in this representation, and thus they
must be found empirically. 

In practice it is a formidable and unsolved problem to extract the constants 
from in flight data for as many as three temporal components to the
transient response behaviour. 
In this paper we describe 
an approximate solution involving a superposition of two exponentials, 
each with illumination-dependent constants. 
This was found to adequately represent 
the detector response on timescales greater than a few seconds, though it
only approximately models the hook response. Our model can be viewed as a 
generalisation of the single exponential model developed for correction
of longer stares by Acosta-Pulido, Gabriel \& Casta\~neda (2000).

In the model considered here, the basic detector response $S$
to an instantaneous step in illumination 
from $S_{\rm \infty p}$ to $S_{\rm \infty}$
is given by the sum of a ``slow'' signal response
$S_{1}$ and a ``fast'' signal response $S_{2}$ as:
\begin{eqnarray}
  S &= &S_{\rm 1} + S_{\rm 2} \\
  S_{\rm 1} &= &(1 - \beta_{\rm 1})\,S_{\rm \infty}\,
(1 - {\rm exp}[-t/\tau_{\rm 1}])\,+\,S_{\rm 01}\,{\rm exp}[-t/\tau_{\rm 1}]\\
  S_{\rm 2} &= &\beta_{\rm 2}\,S_{\rm \infty}\,
(1 - {\rm exp}[-t/\tau_{\rm 2}])\,+\,S_{\rm 02}\,{\rm exp}[-t/\tau_{\rm 2}]
\end{eqnarray}
$S_{\rm 01}$ and $S_{\rm 02}$ are signals corresponding to
the slow and fast response components immediately following 
the illumination step. $S_{\rm 01}$ and $S_{\rm 02}$ are 
related to the corresponding signals immediately prior to the
illumination step, $S_{\rm 1p}$ and $S_{\rm 2p}$, which contain
the information about the illumination history. $S_{\rm 01}$ and $S_{\rm 02}$
are specified by the jump conditions:
\begin{eqnarray}
  S_{\rm 01}&= &\beta_{\rm 1}\,(S_{\rm \infty} - S_{\rm \infty p}) + S_{\rm 1p}\\
  S_{\rm 02}&= &S_{\rm 2p}
\end{eqnarray}
The model is thus specified by four primary parameters: the slow and fast
timescales $\tau_{\rm 1}$ and $\tau_{\rm 2}$, the ``jump''
factor $\beta_{\rm 1}$ and a constant of proportionality
for the fast response $\beta_{\rm 2}$. In order to obtain satisfactory
model fits to complex illumination histories,
each of these primary parameters had to be specified as monotonic
functions of illumination, each with three subsiduary parameters:
\begin{eqnarray}
  \beta_{\rm 1} = \beta_{\rm 10}\,+\,\beta_{\rm 11}\,*\,S_{\rm \infty}^{\beta_{\rm 12}}\\
  \tau_{\rm 1} = \tau_{\rm 10}\,+\,\tau_{\rm 11}\,*\,S_{\rm \infty}^{\tau_{\rm 12}}\\
  \beta_{\rm 2} = \beta_{\rm 20}\,+\,\beta_{\rm 21}\,*\,S_{\rm \infty}^{\beta_{\rm 22}}\\
  \tau_{\rm 2} = \tau_{\rm 20}\,+\,\tau_{\rm 21}\,*\,S_{\rm \infty}^{\tau_{\rm 22}}
\end{eqnarray}
Thus 12 subsiduary parameters in all are needed to characterise the transient
response behaviour of the detectors. In addition, the initial starting 
state of the detector, given by the 
values of $S_{\rm 1p}$ and $S_{\rm 2p}$ immediately prior to
the start of the mapping observation, must be specified.

\subsection{Determination of model parameters}

The parameterisation of the 
illumination dependence of the primary parameters
$\tau_{\rm 1}$, $\beta_{\rm 1}$,
$\tau_{\rm 2}$ and $\beta_{\rm 2}$ was made in 
engineering units (V/s), rather than in astrophysical units such  
as MJy/sr. Thus, it was assumed that there is no dependence of 
the detector response on the wavelength of the IR photons reaching 
the detector. Since the individual pixels of the C100 and the 
C200 arrays behave as independent detectors, each of the 12 
parameters of the detector model had to be separately determined 
for each detector pixel.
 
The ``slow'' primary parameters $\tau_{\rm 1}$ and $\beta_{\rm 1}$ were
found from fits to the detector response to fine calibration
source (FCS) exposures. An example is given in Fig.~3. The model fit
to the signal response on time scales ranging from a few seconds up to the 
duration of the FCS exposure depends mainly on the values of 
$\tau_{\rm 1}$ and $\beta_{\rm 1}$. 
\begin{figure}[ht]
  \begin{center}
\epsfig{file=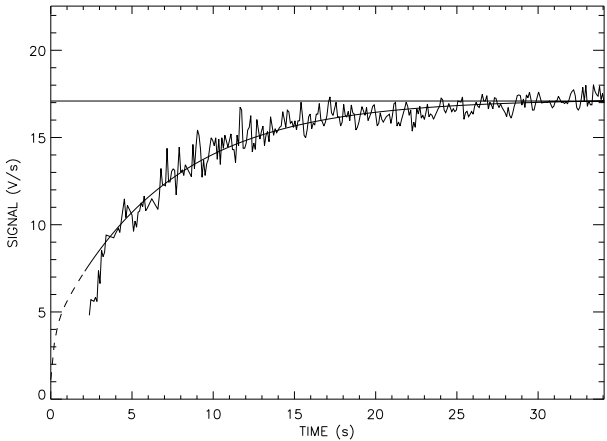, width=8.0cm}
  \end{center}
\caption{Example of a model fit (curve) to the response
of the central pixel of the C100 array to a constant FCS illumination 
starting at $t=0$\,s. Prior to the FCS illumination the detector 
was viewing blank sky. The horizontal line indicates the fitted
value of the illumination. No data was recorded in the first two
seconds following FCS switch-on.
}
\end{figure}
A database of values for
$\tau_{\rm 1}$ and $\beta_{\rm 1}$ was built up for a wide variety
of FCS illuminations, allowing the dependence of $\tau_{\rm 1}$ and 
$\beta_{\rm 1}$ on illumination to be determined according to Eqns.
6$\,\&\,$7. An example showing the illumination dependence of $\beta_{\rm 1}$
for the central pixel of the C100 array is given in Fig.~4. The
solid line is the fitted illumination dependence of $\beta_{\rm 1}$
used to find the values of the parameters $\beta_{\rm 10}$,
$\beta_{\rm 11}$ and $\beta_{\rm 12}$ used in the model. A corresponding
plot for the illumination dependence of $\tau_{\rm 1}$ is given in
Fig.~5.
\begin{figure}[ht]
  \begin{center}
    \epsfig{file=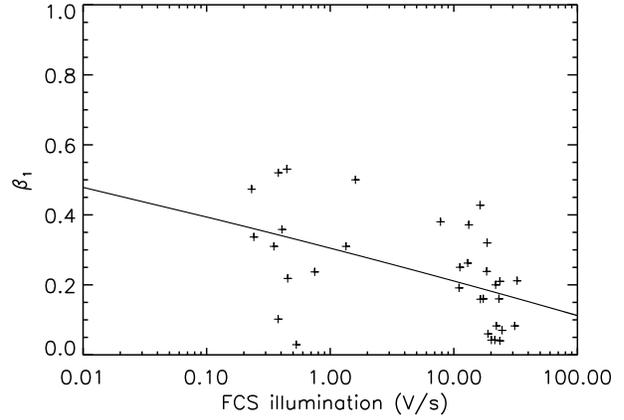, width=8.0cm}
  \end{center}
\caption{Values of the parameter $\beta_{\rm 1}$ for the central pixel
of the C100 array found from fits to the detector response 
to FCS exposures, plotted versus FCS illumination. 
The solid line is a fit of the function 
$\beta_{\rm 1}=\beta_{\rm 10}+\beta_{\rm 11}S_{\rm \infty}^{\beta_{\rm 12}}$ with $\beta_{\rm 10}=2.12$,  $\beta_{\rm 11}=-1.82$,
and $\beta_{\rm 12}=0.022$. These are the parameters adopted for th
model giving the default illumination dependence of $\beta_{\rm 1}$
for this pixel.
}
\end{figure}
\begin{figure}[ht]
  \begin{center}
    \epsfig{file=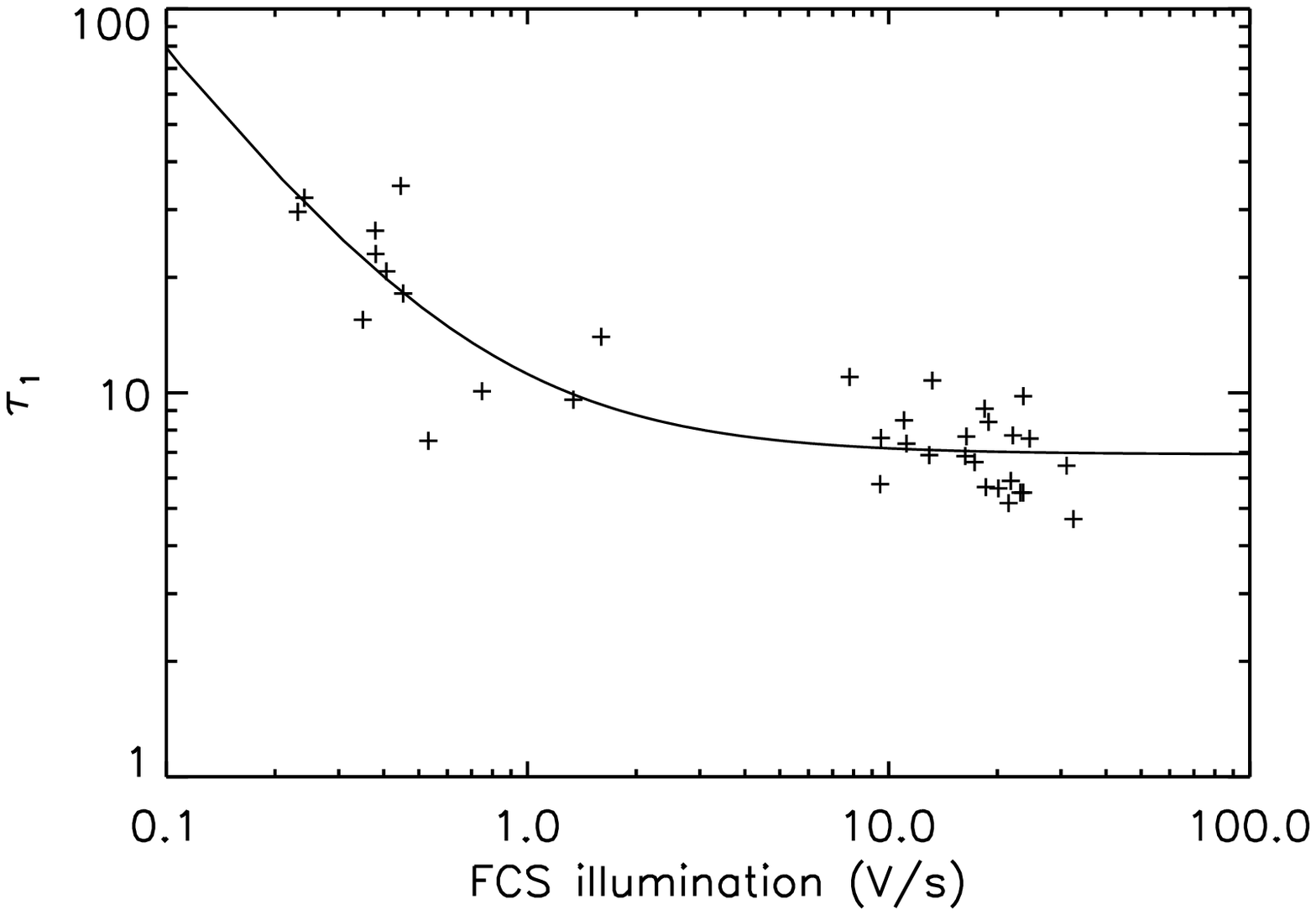, width=8.0cm}
  \end{center}
\caption{As Fig.~4, but for the parameter $\tau_{\rm 1}$.
The solid line is a fit of the function 
$\tau_{\rm 1}=\tau_{\rm 10}+\tau_{\rm 11}S_{\rm \infty}^{\tau_{\rm 12}}$ 
with $\tau_{\rm 10}=6.92s$,  $\tau_{\rm 11}=4.28$,
and $\tau_{\rm 12}=-1.22$. These are the parameters adopted for the
model giving the default illumination dependence of $\tau_{\rm 1}$
for this pixel.
}
\end{figure}
\begin{figure}[ht]
  \begin{center}
    \epsfig{file=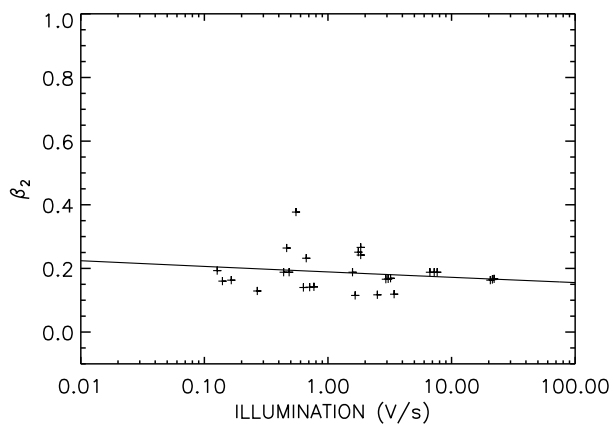, width=8.0cm}
  \end{center}
\caption{Values of the parameter $\beta_{\rm 2}$ for the central pixel
of the C100 array found from self calibration optimisation of
the model on observations of standard point source calibrators,
plotted versus illumination.  
The solid line is a fit of the function 
$\beta_{\rm 2}=\beta_{\rm 20}+\beta_{\rm 21}S_{\rm illum}^{\beta_{\rm 22}}$ with $\beta_{\rm 20}=-0.543$,  $\beta_{\rm 21}=0.723$,
and $\beta_{\rm 22}=-0.0103$, where $S_{\rm illum}$ is the known
illumination of the detector pixel due to the point source calibrator.
These are the parameters adopted for the
model giving the default illumination dependence of $\beta_{\rm 2}$
for this pixel.
}
\end{figure}
\begin{figure}[ht]
  \begin{center}
    \epsfig{file=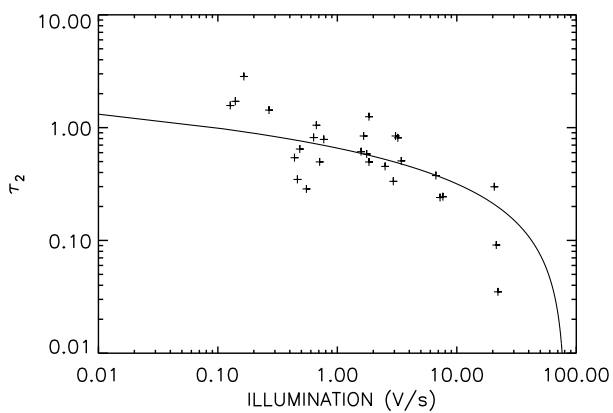, width=8.0cm}
  \end{center}
\caption{As Fig.~4, but for the parameter $\tau_{\rm 2}$.  
The solid line is a fit of the function 
$\tau_{\rm 2}=\tau_{\rm 20}+\tau_{\rm 21}S_{\rm illum}^{\tau_{\rm 22}}$ with $\tau_{\rm 20}=14.89\,s$,  $\tau_{\rm 21}=-14.24$,
and $\tau_{\rm 22}=0.01025$.
These are the parameters adopted for the
model giving the default illumination dependence of $\tau_{\rm 2}$
for this pixel.
}
\end{figure}
At shorter timescales the response is primarily determined by the 
``fast'' primary parameters $\tau_{\rm 2}$ and $\beta_{\rm 2}$. 
However, the FCS exposures could not be used to find $\tau_{\rm 2}$ and 
$\beta_{\rm 2}$ since no data was recorded in the first 
two seconds following
the detector switch-on. This is also the reason why the hook response
is not seen in Fig.~3. Furthermore, there is very probably 
a non-negligible timescale
needed by the FCS to reach its final operating temperature.
Therefore, the ``fast'' parameters (with timescales 
$\sim\,$0.1$\,\le\,\tau\,\le\,\sim\,$5\,s)
were found using a self calibration technique (described in Sect.~4.2),
operating on co-added repeated chopper sweeps crossing standard point source 
calibrators. Observations of standard point source calibrators
of a range of flux densities allowed the dependence of $\tau_{\rm 1}$ and 
$\beta_{\rm 1}$ on illumination to be determined according to the 
parameterisation of Eqns. 8$\,\&\,$9. 
Examples showing the illumination dependence of $\beta_{\rm 2}$
and $\tau_{\rm 2}$ for the central pixel of the C100 array are
given in Figs.~6 and 7, respectively.

\section{THE P32TOOLS ALGORITHM}

In a standard reduction of ISOPHOT data using the interactive analysis 
procedures of PIA, the analysis is made in several irreversible steps,
starting with input ``edited raw data'' at the full time resolution,
and finishing with the final calibrated map. To correct for 
responsivity drift effects, however,
it is necessary to iterate between a sky map and the input data at full
time resolution. This different concept required
a completely new data reduction package. In its development phase, and 
for the determination of the detector model parameters, this 
data reduction package was run as a set of IDL scripts. 
It was later interfaced to PIA as P32TOOLS using a GUI interface 
as described elsewhere in this volume.

There are three basic elements to the P32TOOLS concept. The first step
is signal conditioning, which aims to provide a stream of signal values,
each with an attached time, from which artifacts such as 
glitches, dark current, and non-linearity effects have been removed, but 
which retains the full imprint of the transient response of
the detector to the illumination history at the sharpest available
time resolution. We will refer to such a stream of signals as the
``signal timeline''.  

The second step is the transient correction itself, which is an
iterative process to determine the most likely sky brightness 
distribution giving rise to the observed signals. 
This is a non-linear optimisation problem with the values of
sky brightness as variables. For large maps several hundred independent
variables are involved.

Lastly, there is the calculation  of calibrated transient corrected
maps. This entails the conversion from V/s to MJy/sr, flat fielding,
and the coaddition of data from different detectors pixels. Within
P32TOOLS, this third stage is performed using standard PIA software,
and is therefore not described further here. 

\subsection{Signal Conditioning}

The processing steps for signal conditioning are as follows:

\begin{figure*}[!ht]
  \begin{center}
    \epsfig{file=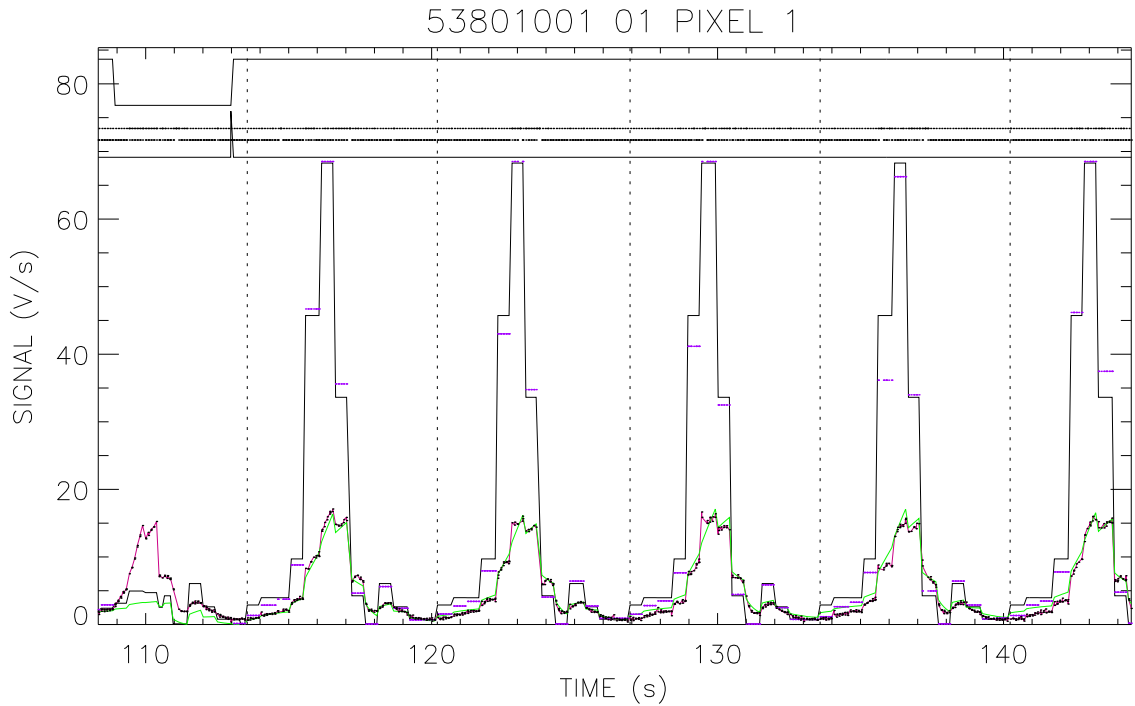, width=15.0cm}
  \end{center}
\caption{Detail from a signal timeline 
from an observation of the standard calibrator Ceres, observed in the
C105 filter. The green line shows the 
fit to the observed signal (shown in red) from pixel 1 of the C100 
array, as found by the P32TOOLS algorithm for 
correction of the transient response behaviour of this pixel. 
5 chopper sweeps, each comprising 13 pointing directions 
(``chopper plateaus'') are shown, divided by vertical dotted lines
separated in time by a duration of 6.1s for each sweep.
The upper line in the figure shows the fine pointing flag;
whereas the first chopper sweep shown was made while the spacecraft was 
slewing between fine pointings (flag value 0), the subsequent
chopper sweeps were made on a single fine pointing (flag value 1).
The remaining horizontal lines denote masks at the full time resolution
of the data denoting glitches, readout status and whether or not a solution
for the illumination could be found by the algorithm. The overall solution
for the sky illumination for this detector pixel is given by the histogram in
black. As described in the text, this overall solution is calculated 
from a combination of individual solutions for illumination on
each successive chopper plateau, which are shown here as purple bars. 
For the very brightest sources, as shown in this example,
the derived illuminations can be up to a factor of 6 brighter than 
the raw data for pixels in the C100 array. In this example the hook 
response is well seen in each chopper sweep at the 9th and 10th chopper 
plateaus.
}
\end{figure*}

\begin{enumerate}

\item Edited Raw Data is imported, after having been corrected in PIA
for the integration ramp non-linearity.

\item The specification of chopper phase in the data is checked
and corrected where necessary.

\item The pointing is established from the instantaneous
satellite pointing information, unlike PIA which uses the commanded
pointing. This also establishes the pointing directions on the slews
between the spacecraft fine pointings. 
Internally defined ``on target flags'' are set for each spacecraft
fine pointing direction, and the central direction and sampling 
intervals directions for the P32 ``natural grid'' (see Sect. 2) 
are established.

\item The integration ramps are differentiated and the data is
corrected for dark current and dependence of photometry 
on reset interval, using standard calibration parameters taken
from PIA. 

\item Systematic variations of the signal according to 
position in the integration ramps are removed from the data. This 
process is called ``signal relativisation'' within P32TOOLS. 
In contrast to the default pipeline or PIA analysis, this allows the use 
of all non-destructive reads, as well as providing a more complete
signal timeline for use in the correction of the transient
behaviour of the detector. The ``signal relativisation'' is 
applied such that signal variations due to the transient 
response behaviour are preserved.

\item Random noise is determined by examining the statistics of the
signals on each chopper plateau. This is necessary due to the
high readout rates of the detectors in many P32 observations. Otherwise,
the determination of noise from individual readout ramps, as done in a 
standard PIA analysis, would be insufficiently precise.

\item Spikes are detected and removed from the data in a first-stage
deglitching procedure. This operates at the full time resolution of the data
given by the non-destructive read interval.

\item A second stage deglitching procedure is performed,
operating at the time resolution of the chopper plateau 
(which comprise of at least 16 non-destructive read intervals).
This procedure (described by Peschke \& Tuffs in this volume)
removes any long lived glitch ``tails''
following the spikes detected in the first-stage deglitch.

\end{enumerate}

\subsection{Transient Correction}

The calculation to correct for the transient response behaviour
is made separately for each detector
pixel. The kernel of the procedure is to
solve for the illumination corresponding to the
signal on each chopper plateau. This is done in a binary chop scheme
for set values of the 12 detector parameters. The values for $S_{\rm 1p}$
and $S_{\rm 2p}$ needed to solve for illumination on each chopper plateau
are calculated from the response of the model to the preceding illumination
history.
\begin{figure*}[!ht]
  \begin{center}
    \epsfig{file=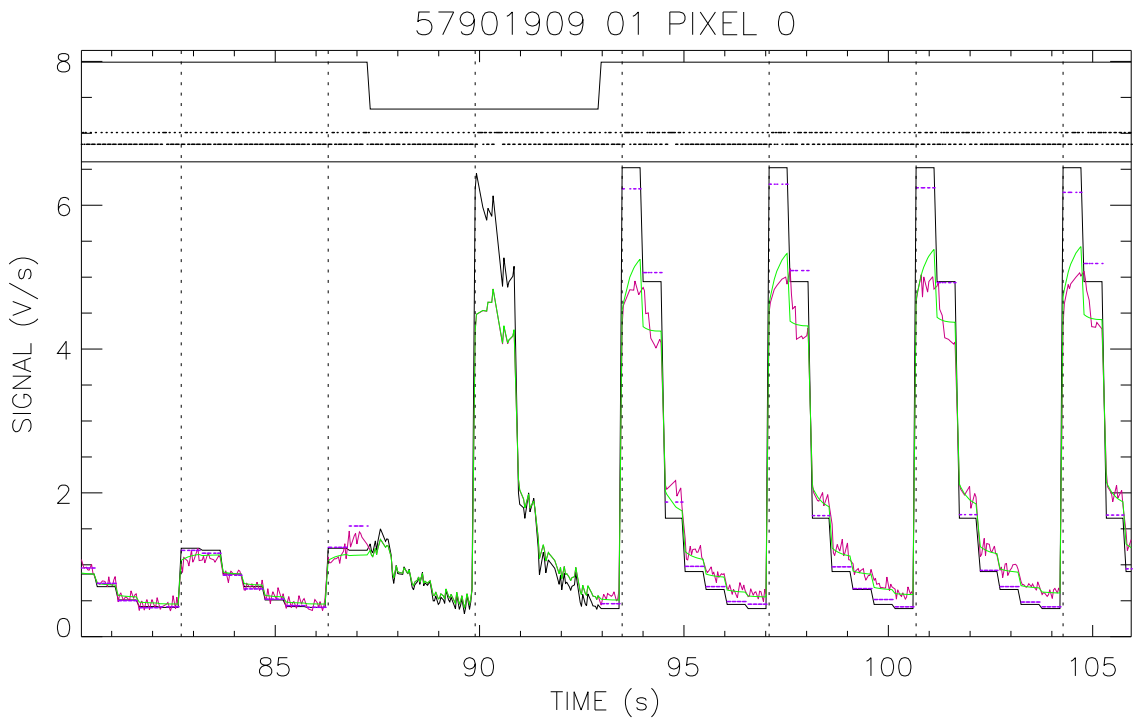, width=15.0cm}
  \end{center}
\caption{Detail from a timeline of signal versus time
from another observation of Ceres, this time observed in the
C160 filter using the C200 detector array. The key to the plot is as given 
in the caption to Fig.~8. This example shows the transition from 
a spacecraft fine pointing for which the chopper is sweeping
through beam sidelobes to a fine pointing where the chopper samples 
the beam kernel. In this case the red line corresponding to the data is
not seen during the intervening spacecraft slew, as it is
exactly overplotted by the green line showing the fit to the data
(see text). The chopper sweeps on the second fine pointing encompassing the
beam kernel show a typical enhancement of the source-background contrast 
induced by the algorithm for correction of the transient response.
}
\end{figure*}
After solving for the illumination on the current chopper
plateau, the found value is combined with previous solutions
encountered for the same sky position from previous chopper
plateaus viewing the same sky direction. 
The response of the
detector pixel to the illumination history is recalculated and
a solution is found for the illumination viewed on the next chopper 
plateau. This procedure is continued until all the signal timeline
has been processed. Then a second pass through the data
can be made, and so on, until the overall solution 
for the sky brightness distribution no longer changes.
The goodness of the solution is quantified through 
a value for $\chi^{\rm 2}$,
calculated from a comparison between the model fit and the signal timeline.
The random uncertainties assigned to each signal sample in the timeline
can optionally be recalculated from the self consistency of 
solutions for the illumination found in a given sky direction.
Examples of model fits to the signal timeline are shown in
Fig.~8 for the C100 detector and in Fig.~9 for the C200 detector.

The algorithm can also be used for datasets for which the 
readout rate was too rapid to allow all the data to be transmitted 
to the ground. In such cases the illumination history can still be
determined at all times on fine pointings, 
even where there are gaps in the signal timeline. On slews,
the illumination history during the gaps are either interpolated 
in time, or taken from the nearest previously solved direction
on the P32 natural grid. 

For most observations using the C200 detector, and observations
of moderate intensity sources made using the C100 detector, 
good fits to the entire unbroken signal timeline
can be obtained for each detector pixel. For observations of
bright sources using the C100 detector, however, 
it is often necessary to break up
the solutions into smaller segments, according to the distribution
of the sources in the field mapped. This is best achieved by 
solving separately for the portions of the signal timeline
passing through the bright sources. The most conservative
processing strategy, appropriate for observations using the C100
detector of complex brightness distributions with high source to 
background contrast, is to break up the calculation into 
time intervals corresponding to individual spacecraft fine pointings. 
This means that no processing of the slews is required.

Several processing options are available, to optimise solutions
for the particular characteristics of each dataset:

\bigskip

\noindent {\it Determination of detector starting state}

\medskip

The default operation of the program is to assume that the
detector is in equilibrium at the start of the observation,
viewing the sky direction corresponding to the first detector plateau.
This corresponds to the case $S_{\rm 1p}\,=\,S_{\rm 01}\,=\, S_{\rm \infty}$
and $S_{\rm 1p}\,=\,0$.
This will very rarely be the case, however, since an FCS measurement 
which may not be perfectly matched to the sky brightness will have been
made just seconds prior to the start of the spacecraft raster.
\begin{figure}[h]
  \begin{center}
    \epsfig{file=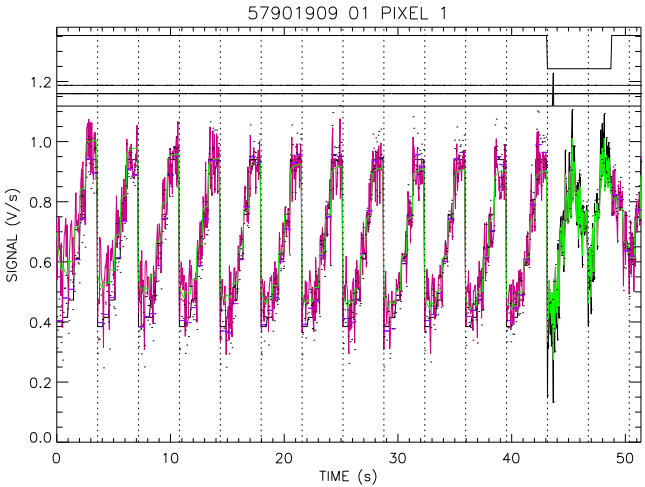, width=8.5cm}
  \end{center}
\caption{A further detail from the timeline of signal versus
time for the observation of Ceres in the C160 filter. This time
the fit to the data for the very first fine pointing is shown, to
illustrate the fit to the switch-on transient behaviour of the detector
(see Sect.~4.2) For a constant repeated illumination pattern in
each chopper sweep, both the data (red) and the fit (green) gradually
move lower through the first half minute of the observation.
}
\end{figure}
Therefore, the program can optionally find a solution for the
detector starting state, as characterised by the values of
$S_{\rm 1p}$ and $S_{\rm 2p}$ immediately prior to the
start of the first chopper plateau. This is done by solving
for $S_{\rm 1p}$ and $S_{\rm 2p}$ for a fixed $S_{\rm \infty}$
on the first plateau, where $S_{\rm \infty}$ is determined from
the last chopper plateau in the first spacecraft pointing to view
the same sky direction. The timeline for the first spacecraft
pointing is then repeatedly processed until a stable solution for
$S_{\rm 1p}$ and $S_{\rm 2p}$ prior to the first plateau is found.
This option was found particularly useful for
data taken with the C200 detector. An example of a model fit to a
timeline where a solution for the detector starting states was found 
in this way is shown in Fig.~10. 

\bigskip

\noindent {\it ``Self calibration'' of detector parameters}

\medskip

In order to determine the primary detector parameters $\beta_{\rm 2}$
and $\tau_{\rm 2}$ in the first instance, 
``a self calibration'' processing technique was used
by which any of the 12 detector parameters (Sect.~3.1) could be determined
from observations of bright sources. Sources of any arbitrary
brightness distribution and position can be used. However, to
determine the detector parameters for the 
``fast'' response of the detector, point source calibrators of
known brightness were used, and the solutions for $\tau_{\rm 2}$ and
$\beta_{\rm 2}$ were additionally constrained by the requirement that the
solutions for illumination were consistent with the known flux densities 
of the calibrators.

The self calibration procedure works by simply repeating the
optimisation process for different subsets of the 12 detector
parameters, specified by the user. Each combination
of detector parameters produces an individual solution for the sky brightness 
distribution, as well as a value for $\chi^{\rm 2}$. A search is made 
in the parameter space of the detector parameters
until a minimum is found in $\chi^{\rm 2}$. This is a lengthy
process, however, making it advisable to perform self calibrations
on limited portions of the timeline. This is often chosen to correspond
to a single spacecraft pointing where the chopper sweeps for the detector
pixel being investigated pass through bright structure. The process
can be further accelerated in cases that the repeated chopper sweeps
through a source give the same repeated signal pattern. Then, 
the optimisation can be performed on an average of the coadded chopper 
sweeps performed on a single spacecraft pointing. This 
latter technique is referred to in P32TOOLS as a ``composite 
self calibration''. Some examples of the useage of the self calibration
procedure are given in this volume by Schulz et al. (2002).

\bigskip

\noindent {\it Processing of slews}

\medskip

The calculation of the illumination history for data taken
during the slews is less straightforward than for data 
taken during the spacecraft fine pointings. This is because
the instantaneous pointing directions during slews are effectively 
at random positions with respect to the P32 natural grid. Therefore,
the use of solutions for the sky brightness distribution
on the P32 natural grid found from analysis of previous fine pointings
will in general lead to inaccuracies in the determination of
the illumination history during the slews. 
There are two main approaches which can be adopted to counter this 
problem.

The first is to solve for the illumination not over a chopper
plateau, but for individual readouts in the signal timeline
during the slew. This automatically provides a solution in which
the model fit exactly matches the data, as there is only one
fitted data point for each determined illumination. An example
is given in the fit to the signal timeline for the C200 detector
shown in Fig~9. This approach is useful for observations of bright
structured sources in which the variations in illumination
with position on the slew on timescales of a chopper plateau
exceed the signal to noise on individual data samples. It is
the default processing option for the C200 detector.

In cases of low signal to noise, and for all instances where
not all data are transmitted to the ground, solving
for the average illumination on each chopper plateau is
a preferable technique. Otherwise, the values of $S_{\rm 1p}$
and $S_{\rm 2p}$ calculated just prior to the first chopper plateau
on the next fine pointing can be wildly inaccurate, which can corrupt
the solutions for the remainder of the signal timeline of the observation.
Processing on a plateau by plateau basis is the default 
processing option to calculate the illumination history during
slews for observations with the C100 detector. 

\section{PHOTOMETRIC PERFORMANCE}

In general the corrections in integrated flux densities
made by the algorithm depend on the source brightness, structure,
the source/background ratio, and the dwell time on each chopper plateau.
The largest corrections are for bright point sources on faint backgrounds.

Here we give as an example 
results achieved for the faint
standard star HR~1654. This source was not used in the determination of the 
detector model parameters, so it constitutes a test of the photometric
performance of ISOPHOT in its P32 observing mode. The derived integrated flux 
densities, with and without correction for the transient response behaviour of 
the C100 detector, were compared in Fig.~11 with predicted flux densities from
a stellar model. The corrected photometry is in reasonable agreement with 
the theoretical predictions. As expected, there is a trend for observations
with larger detector illuminations 
to have larger corrections in integrated photometry. 


\begin{figure}[h]
  \begin{center}
    \epsfig{file=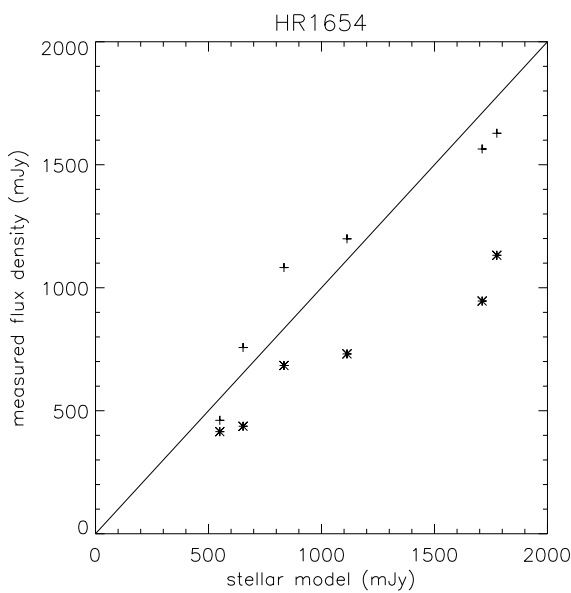, width=7.5cm}
  \end{center}
\caption{
The measured integrated flux densities of the faint standard star HR~1654
plotted against the predicted flux densities from a stellar model. The
observations were done in various filters using the C100 detector. The
photometry derived with and without processing with P32TOOLS is shown with
stars and crosses, respectively.
}
\end{figure}
A good linear correlation is also seen between integrated flux densities
of Virgo cluster galaxies (Tuffs et al. 2002), derived from P32 ISOPHOT 
observations processed using the P32TOOLS algorithm, and flux densities from 
the IRAS survey (Fig.~12). The interpretation of these measurements
constitutes  the first science application (Popescu et al. 2002) of P32TOOLS 
algorithm. The ISOPHOT observations of Virgo cluster galaxies 
were furthermore used to derive the ratios of fluxes measured by ISO to those 
measured by IRAS. The ISO/IRAS ratios were found to be
0.95 and 0.82 at 60 and 100$\,\rm \mu m$, respectively, after scaling the 
ISOPHOT measurements onto the COBE-DIRBE flux scale (Tuffs et al. 2002).
\begin{figure}[h]
  \begin{center}
    \epsfig{file=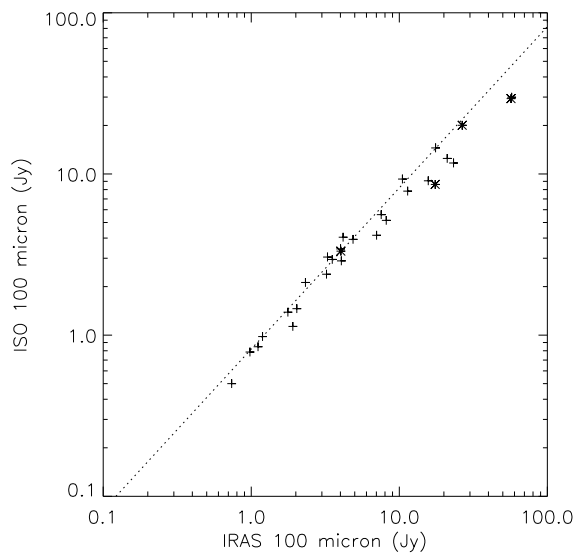, width=7.5cm}
  \end{center}
\caption{Integrated and colour-corrected flux densities of
Virgo cluster galaxies measured in the ISO C100 filter versus
the corresponding flux densities measured by IRAS in its 
100$\,\rm \mu m$ band (taken from Fig.~7 of Tuffs et al. 2002). The dotted
line represents the relation ISO/IRAS=0.82. 
}
\end{figure}
\begin{figure}[h]
  \begin{center}
    \epsfig{file=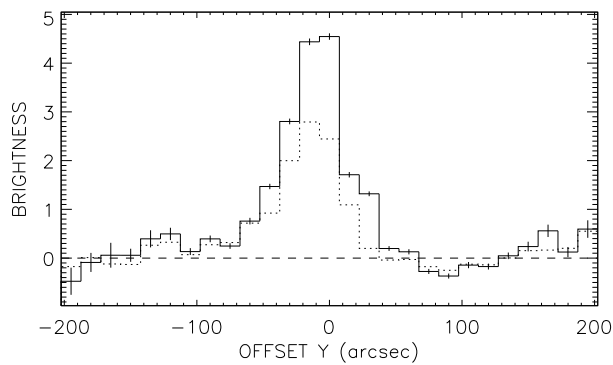, width=8.0cm}
  \end{center}
\caption{Brightness profiles along the Y spacecraft direction 
through the standard star HR~1654
at 100$\,\rm \mu m$. The solid line represents the brightness profile
obtained after processing with P32TOOLS, while the dotted line 
shows the profile derived from identically 
processed data, but without correction for the transient response of
the detector.
}
\end{figure}

Fig.~13 shows an example of a brightness profile through HR~1654 at
100$\,\rm \mu m$, for data processed with and without the responsivity 
correction. Some 95$\%$ of the flux density has been recovered by P32TOOLS. 
Without the correction, some 30$\%$ of the integrated emission is missing
and the signal only reaches 50$\%$ of the peak illumination.
The local minimum near 60~arcsec in the Y offset is a typical hook
response artifact, where the algorithm has overshot the true solution.
This happens for rapid chopper sweeps passing through the beam kernel. 
This is a 
fundamental limitation of the detector model, which, as described in 
Sect.~3.1, does not correctly reproduce the hook response on timescales
of up to a few seconds. This problem is particularly apparent for
downwards illumination steps. The only effective antidote is to
mask the solution immediately following a transition through a bright
source peak. Another effect of the inability to model the
hook response is that the beam profile becomes somewhat distorted.
This also has the consequence, that for observations of bright sources,
the measured FWHM can become narrower than for the true point spread function,
as predicted from the telescope optics and pixel footprint. An example
of an extremely bright point source showing this effect is 
given in Figs.~14 \& 15, depicting maps of Ceres in the C105 
filter, respectively made without and with the correction for the transient 
response of the C100 detector. 
\begin{figure}[!ht]
  \begin{center}
    \epsfig{file=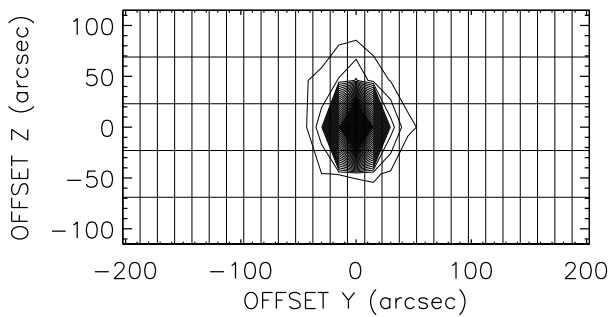, width=8.0cm}
  \end{center}
\caption{Contour map of Ceres in the C105 filter after
responsivity drift correction. 50 linear
contours have been plotted between 33 and
2519 MJy/sr. Measured integrated flux density
after background subtraction is $98.4$\,Jy\,$\pm\,0.2$\,Jy (random) 
$\,\pm\,6\%$ (systematic). The actual flux density of this standard
calibrator (from a stellar model) is 109\,Jy.
}
\end{figure}
\begin{figure}[h]
  \begin{center}
    \epsfig{file=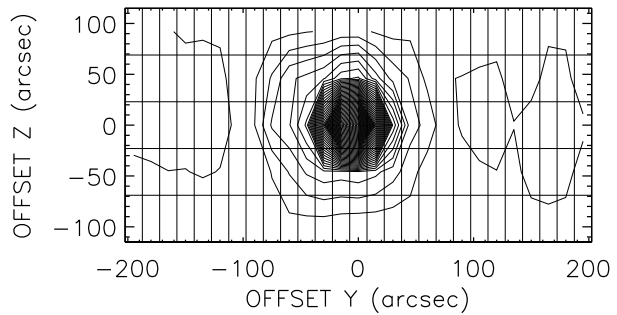, width=8.0cm}
  \end{center}
\caption{Contour map of Ceres in the C105 filter without
correction for the transient response behaviour of the detector.
50 linear contours have been plotted between 11.5 and
334 MJy/sr. Measured integrated flux density
after background subtraction is $28.0$\,Jy\,$\pm\,0.07$\,Jy (random) 
$\,\pm\,4\%$ (systematic).
}
\end{figure}

Despite the limitations due to the lack of a proper modelling of the
hook response, the algorithm can effectively correct for artifacts 
associated with the transient response on timescales from a few seconds
to a few minutes. This is illustrated in Figs.\,16\,\&\,17 
by the maps of the interacting
galaxy pair KPG~347 in the C200 filter 
(again, after and before correction for the transient response
of the detector, respectively). 
The uncorrected map shows a spurious elongation in
the direction of the spacecraft Y coordinate,
which is almost completely absent in the corrected map. If uncorrected,
such artifacts could lead to false conclusions about the brightness
of FIR emission in the outer regions of resolved sources. Also visible in
the corrected map is a trace of a beam sidelobe.

\begin{figure}[h]
  \begin{center}
    \epsfig{file=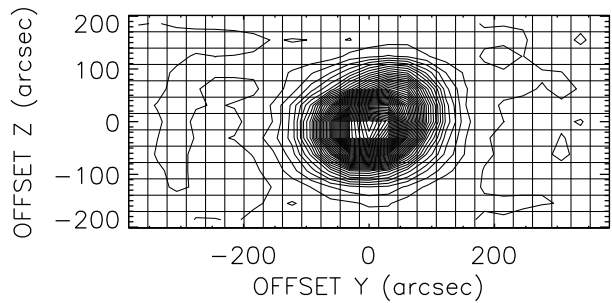, width=8.0cm}
  \end{center}
\caption{The interacting galaxy pair kpg~347 observed in the
C200 filter after processing with P32TOOLS.
}
\end{figure}
\begin{figure}[h]
  \begin{center}
    \epsfig{file=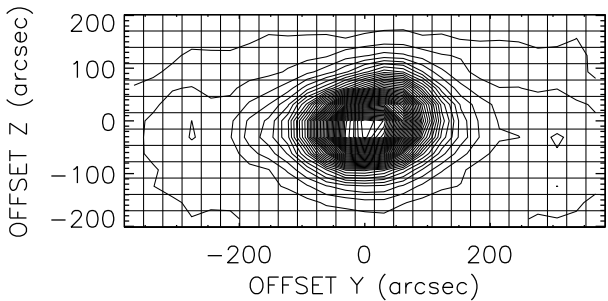, width=8.0cm}
  \end{center}
\caption{The interacting galaxy pair kpg~347 with identical
processing, except that the correction for the transient 
response behaviour of the detector has been omitted
}
\end{figure}

The image of M~101 in Fig.~18, made using the C100 detector with
the C100 filter, is given as a state of the art 
example of what can be achieved with a careful interactive 
processing of P32 data using P32TOOLS. In addition to the
transient response corrections and a masking of residual
hook response artifacts, a time dependent flat field 
has been applied. The spiral structure of the galaxy, with
embedded HII region complexes and a component of diffuse
interarm emission can clearly be seen.

\begin{acknowledgements}

This work was supported by grant 50-QI-9201 of the Deutsches Zentrum f\"ur
Luft- und Raumfahrt. I would like to thank all those who have helped 
me in many ways in the development of the algorithm described here. 
Richard Tuffs would like to thank his colleagues at the 
Max-Planck-Institut f\"ur Kernphysik, in particular 
Prof. Heinrich V\"olk, for their support and encouragement. 
We have also benefited from many useful discussions with 
Drs. R. Laureijs, S. Peschke and B. Schulz and the team in the 
ISO data centre at Villafranca, with Prof. D. Lemke and Dr. U. Klaas at the
ISOPHOT data centre at the Max-Planck-Institut f\"ur Astronomie,
and with Drs. N. Lu and I. Khan at the Infrared Processing
and Analysis Center. 

\end{acknowledgements}

\end{document}